\documentclass[preprint,12pt]{elsarticle}
\usepackage{graphicx}
\usepackage{amsfonts}
\usepackage{amsmath}
\usepackage{url}
\usepackage{epstopdf}
\usepackage{hyperref} 
\usepackage{color}

\def\ve{\varepsilon}
\def\C{{\mathcal C}}

\def\Tr{\mathrm{Tr}}
\def\R{{\mathbb R}}

\def\CC{{\mathbb C}}

\def\det{\mathrm{det}}

\journal{Physica A}

\begin{document}

\begin{frontmatter}
\title{Emergence of correlations between securities at short time scales}

\author{Sebastien Valeyre}
\address{
John Locke Investment, 38 Avenue Franklin Roosevelt, 77210 Fontainebleau-Avon, France}
\ead{sebastien.valeyre@jl-investments.com}

\author{Denis~S.~Grebenkov}
\address{
Laboratoire de Physique de la Mati\`{e}re Condens\'{e}e, \\ 
CNRS -- Ecole Polytechnique, 91128 Palaiseau, France}
\ead{denis.grebenkov@polytechnique.edu}

\author{Sofiane Aboura}
\address{
Universit\'e de Paris XIII, Sorbonne Paris Cit\'e, 93430 Villetaneuse, France}
\ead{sofiane.aboura@univ-paris13.fr}

\date{\today}

\begin{abstract}
The correlation matrix is the key element in optimal portfolio
allocation and risk management.  In particular, the eigenvectors of
the correlation matrix corresponding to large eigenvalues can be used
to identify the market mode, sectors and style factors.  We
investigate how these eigenvalues depend on the time scale of
securities returns in the U.S. market.  For this purpose, one-minute
returns of the largest 533 U.S. stocks are aggregated at different
time scales and used to estimate the correlation matrix and its
spectral properties.  We propose a simple lead-lag factor model to
capture and reproduce the observed time-scale dependence of
eigenvalues.  We reveal the emergence of several dominant eigenvalues
as the time scale increases.  This important finding evidences that
the underlying economic and financial mechanisms determining the
correlation structure of securities depend as well on time scales.
\end{abstract}

\end{frontmatter}

\section{Introduction}

How do the eigenvalues of securities correlation matrices emerge at
different time scales?  This fundamental question is important because
cross-correlations change over different investment horizons while a
reliable empirical determination of the correlation matrix remains
difficult due to its time and frequency dependence.  This was first
evidenced by Epps, who demonstrated the decay of correlations among
U.S. stocks when shifting from daily to intra-daily time scales (or
frequencies) \cite{Epps79}.  In other words, the price correlation
decreases with the duration of the time interval over which price
changes are measured.  The economic argument behind the Epps effect is
that the information is not instantaneously transmitted at shorter
time intervals, where the average adjustment lag in response of prices
lies approximately between 10 and 60 minutes.  This appears to reduce
the scope of the Efficient Market Hypothesis \cite{Fama70} at short
time scales given that tick data prices seem to adjust to new
information only after a lag time, thus do not reflect all available
information.  Since its inception, the Epps effect has been confirmed
by several studies, although its impact has been progressively
declined in the NYSE, indicating that the market becomes increasingly
more efficient
\cite{Toth06}.

The dependence of securities cross-correlations on time scales can be
captured via the eigenvalues of the correlation matrix.  In
particular, the largest eigenvalue reflects changes in the average
correlation between stocks, whereas the corresponding eigenvector is
associated to the ``market mode''.  Kwapien {\it et al.} showed a
significant elevation of the largest eigenvalue with increasing time
scale using data from 1 minute to 2 days from NYSE, NASDAQ and
Deutsche B\"orse (1997-1999) \cite{Kwapien04}.  Using high-frequency
stock returns from NYSE, AMEX and NASDAQ (1994-1997), Plerou {\it et
al.} supported the idea that the largest eigenvalue and its
eigenvector reflect the collective response of the entire market to
stimuli such as certain news breaks (e.g., central bank interest rates
hikes) \cite{Plerou02}.  This is particularly true during periods of
high volatility when the collective behavior is enhanced.  Coronnello
{\it et al.}  confirmed that the largest eigenvalue, computed from
5-minute data, describes the common behavior of the stocks composing
the LSE stock index (2002) \cite{Coronnello05}.

As firms having similar business activities are correlated, some other
eigenvectors can economically be interpreted as business sectors
\cite{Gopikrishnan01}.  So, Gopikrishnan {\it et al.} computed the
eigenvectors of cross-correlation matrices of 1000 U.S. stocks at a
30-minute scale (1994-1995) and a 1-day scale (1962-1996)
\cite{Gopikrishnan01}.  They found that the correlations in a business
sector, captured via an eigenvector, were stable in time and could be
used for the construction of optimal portfolios with a stable Sharpe
ratio.  In the same vein, as similar trading strategies induce
cross-correlations in stocks, some eigenvectors can be financially
interpreted as style factors.  The corresponding eigenvalues are thus
expected to exhibit non-trivial dependence on time scales.  However,
an accurate statistical analysis of multiple eigenvalues at different
time scales is challenging due to measurement noises.  In fact, as the
correlation matrix is estimated from time series of stocks' returns,
its elements are unavoidably random and thus prone to fluctuations.
These fluctuations become larger as the length of time series is
reduced, i.e., when the time scale is increased.  While the largest
eigenvalue typically exceeds the level of fluctuations by two orders
of magnitude, the other eigenvalues rapidly reach this level and
become non-informative.  Several researchers employed the random
matrix theory to distinguish economically significant eigenvalues from
noise \cite{Laloux99,Pafka04,Burda04,Potters05,Conlon07}.  In
particular, Laloux {\it et al.} showed that only 6\% of the
eigenvalues carried some information of the S\&P 500 (1991-1996),
while the remaining 94\% eigenvalues were hidden by noise
\cite{Laloux99}.  Guhr and Kalber proposed an alternative statistical
approach to reduce noise that they called ``power mapping''
\cite{Guhr03}.  Andersson {\it et al.}  extended this work by comparing
the power mapping approach to a standard filtering method discarding
noisy eigenvalues for Markowitz portfolio optimization using daily
Swedish stock market returns (1999-2003) \cite{Andersson05}.

In this paper, we consider the correlation matrix of financial
securities and investigate the emergence of its eigenvalues at small
time scales.  As the financial literature on this critical issue
remains sparse, this research fills the gap by investigating the
eigenvalues at intraday time scales using 1-min returns.  We propose a
simple model, coined the ``lead-lag factor model'', as an adaptation
of the well-known ``one-factor marker model'' \cite{Ross76} to smaller
time scales and to multiple sectors and style factors.  In this model,
stock returns are correlated to the returns of selected factors at
earlier time steps.  A detailed description of the eigenvalues as
functions of the time scale is then derived.  An empirical validation
is performed on long time series of 1-min returns of a large universe
of U.S. stocks.  To get several significant eigenvalues at time scales
from 1 minute to 2 hours, the correlation matrix was estimated over
the whole available period (2013-2017) so that variations of
cross-correlations over time were ignored (note that the dynamics of
the eigenvalues and eigenvectors over time has been investigated
elsewhere \cite{Conlon09b,Allez12,Buccheri13}).  In spite of its
simple character, the lead-lag factor model is shown to be able to
reproduce the dependence of large eigenvalues on the time scale.

The paper is organized as follows.  In Sec. \ref{sec:results}, we
estimate the correlation matrix of U.S. stocks' returns at different
time scales and present the empirical dependence of large eigenvalues
on the time scale.  To rationalize the observed behavior, we develop
in Sec. \ref{sec:theory} the lead-lag factor model and compare it to
empirical results.  Section \ref{sec:conclusion} summarizes and
concludes.  Some derivations and more technical analysis of the
lead-lag factor model are presented in Appendices.

\section{Empirical results}
\label{sec:results}

\subsection{Data description}

We study the correlation structure of a universe
%managed by John Locke Investment 
that includes 533 U.S. stocks whose capitalization exceeded 1 billion
dollars in 2013.  For the considered period from 1st of January 2013
to 28th of June 2017, our database contains 338~176 1-min returns for
each stock.  We have also verified that the arithmetic aggregation of
returns, $r_i(1) + \ldots + r_i(\tau)$, is almost identical to
considering the product $(1+r_i(1))\ldots (1+r_i(\tau)) - 1$, given
that the 1-min returns $r_i(t)$ are very small.

From the time series of 1-min returns, we estimate the correlation
matrix over the whole available period, and then compute its
eigenvalues.  Then we aggregate the returns into 2-min, 4-min, ...,
128-min returns, producing time series with 169~088, 84~544, ...,
2~642 points, respectively.  At each time scale $\tau$, we repeat the
computation to investigate the dependence of the eigenvalues on
$\tau$.

\subsection{Empirical results}

\begin{figure}
\begin{center}
\includegraphics[width=67.5mm]{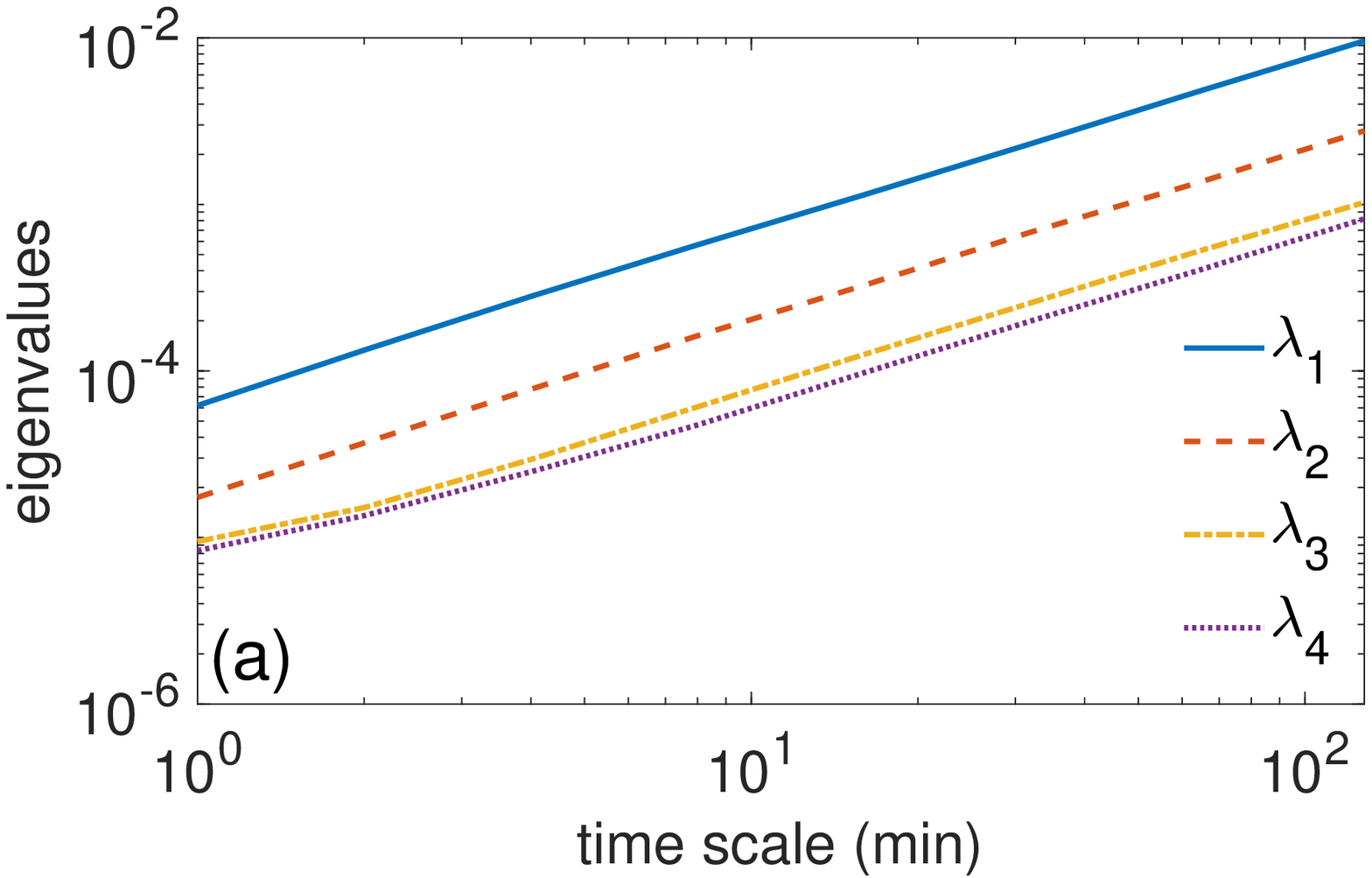} % returns1min_eigen_C_USAnew.eps}
\includegraphics[width=67.5mm]{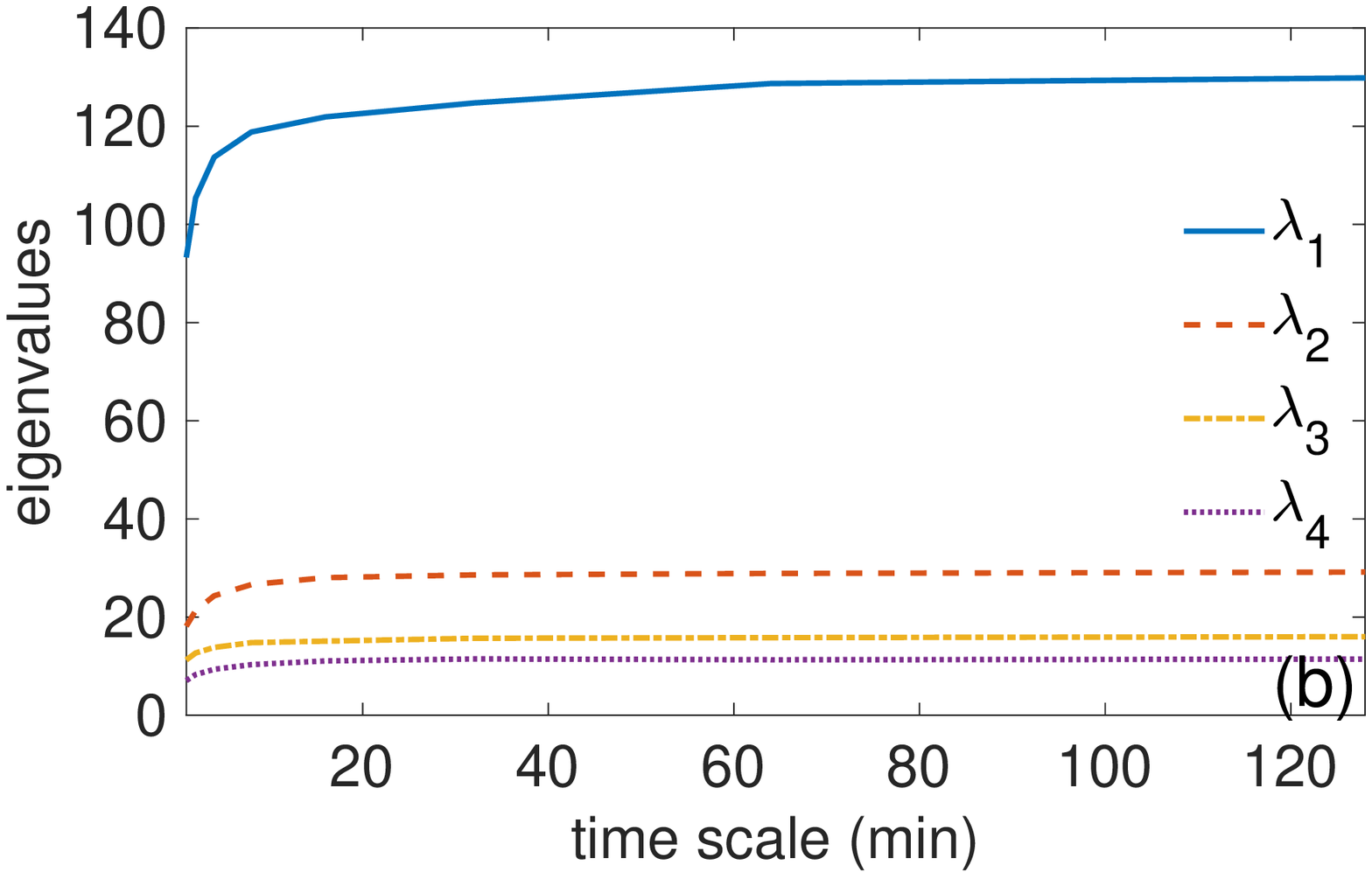} % returns1min_eigen_rho_USAnew.eps}
\end{center}
\caption{
Four largest eigenvalues of the covariance matrix {\bf (a)} and of the
correlation matrix {\bf (b)} for returns of 533 U.S. stocks, computed
by aggregating 1-min returns with the time scale $\tau$, varying from
1 minute to 128 minutes (2 hours). }
\label{fig:returns1min_eigen_USA}
% [LamC, Lamrho] = JLI_intraday_create_C_fig4(LamC, Lamrho);
\end{figure}

Figure \ref{fig:returns1min_eigen_USA}a shows the four largest
eigenvalues of the covariance matrix of 533 U.S. stocks' returns,
computed by aggregating 1-min returns with the time scale $\tau$,
ranging from 1 minutes to 128 minutes (2 hours).  The first two
eigenvalues exhibit almost linear growth with $\tau$, the others show
minor deviations from linearity at small $\tau$ but scale linearly
with $\tau$ at large $\tau$.  This behavior reflects the
diffusion-like growth of the variance of aggregated returns; in
particular, if the returns were independent, the eigenvalues of the
corresponding covariance matrix, $C_{ij} = \tau
\sigma_i^2 \delta_{ij}$, would be just $\lambda_i = \tau \sigma_i^2$,
and thus proportional to $\tau$.  Although correlations affect this
linear growth, their effect is subdominant, at least for large
eigenvalues, as witnessed by Fig. \ref{fig:returns1min_eigen_USA}a.
To highlight the effect of correlations, we focus on the eigenvalues
of the {\it correlation} matrix.  This choice is also justified from
the financial point of view to level off the variability of stocks
volatilities.

Figure \ref{fig:returns1min_eigen_USA}b shows the four largest
eigenvalues of the correlation matrix of the same 533 U.S. stocks'
returns.  If the returns were independent, the correlation matrix
would be the identity, and thus all its eigenvalues would be equal to
$1$.  The growth of these eigenvalues with the time scale $\tau$
indicates strong cross-correlations between stocks.  The largest
eigenvalue can be naturally attributed to the market mode, whereas the
next eigenvalues correspond to different sectors and style factors.

After a sharp growth at short time scales (few minutes), the
eigenvalues slowly approach to their long-time limits.  The existence
of these upper bounds is expected because the sum of eigenvalues of a
correlation matrix is equal to its size (i.e., to the number of
stocks, $N$).  This saturation effect contrasts with the unlimited
growth of eigenvalues of the covariance matrix
(Fig. \ref{fig:returns1min_eigen_USA}a).  Finding the functional form
of this approach and identifying its characteristic time scales
present the main aim of our work.  Recently, Benzaquen {\it et al.}
proposed a multivariate linear propagator model for dissecting
cross-impact on stock markets and revealing their dynamics
\cite{Benzaquen17}.  Due to its very general form accounting for
both cross-correlations and auto-correlations of stocks, the proposed
model contains too many parameters, while the resulting formulas are
not explicit.  Our ambition is rather the opposite and consists in
suggesting an explicit model, as simple as possible, that would
capture the empirical results shown in
Fig. \ref{fig:returns1min_eigen_USA}b and thus provide a minimalistic
framework for their financial interpretation.

\section{The lead-lag factor model}
\label{sec:theory}

\subsection{Basic lead-lag one-factor model}
\label{sec:one-factor}

We consider a trading universe with $N$ assets.  In a conventional
one-factor model, the return of the $i$-th asset at time $t$,
$r_i(t)$, is modeled as a combination of a specific, asset-dependent
random fluctuation, $\ve_i(t)$, and an overall market contribution,
$R(t)$,
\begin{equation}
r_i(t) = \ve_i(t) + \beta R(t) ,
\end{equation}
with a market sensitivity $\beta$ (that we generalize below to other
factors).  The asset-specific random fluctuations $\ve_i(t)$ are
typically modeled as independent centered Gaussian variables with
volatilities $\sigma_i$.

We propose a modification of this conventional model by incorporating
the lead-lag effect, in which the $i$-th asset return at time $t$ is
influenced by a common factor $R(t-k)$ at earlier times $t-k$, with
progressively decaying weights:
\begin{equation}  \label{eq:ri}
r_i(t) = \ve_i(t) + \beta \sum\limits_{k=0}^\infty \alpha^k R(t-k)  ,
\end{equation}
where $0 \leq \alpha < 1$ characterizes the relaxation time of the
memory decay.  Note that the upper limit of the sum in
Eq. (\ref{eq:ri}) is formally extended to infinity, bearing in mind
that contributions for very large $k$ are exponentially small.  We
will analyze the model in the stationary regime as $t\to\infty$ in
order to eliminate transient effects.

The common term $R(t)$ can be interpreted as an idealized factor
without auto-correlations in an efficient market that most stocks
follow with a lead lag delay.  We model therefore $R(t)$ by
independent centered Gaussian variables with volatility $\Sigma$.  The
term $R(t)$ can represent the market mode but also sectors or style
factors, or any popular trading portfolio.  Moreover, $R(t)$ can also
be interpreted as being linked to the market order transactions for a
particular strategy (market, sector or styles).  In this light, our
model can be seen as an extension of the Kyle model \cite{Kyle85} that
explains the impact of transactions on price for a single stock and
without delay.  Here, we consider multiple stocks and include an
exponential decay of the impact.  While more sophisticated models with
a power law decay of the impact were proposed
\cite{Benzaquen17,Bouchaud09}, we will show that our minimalistic model
is enough to reproduce a slow growth of the eigenvalues of the
correlation matrix.  For the sake of clarify, we first analyze this
basic lead-lag one-factor model and then discuss its several
straightforward extensions.

The one-factor relation (\ref{eq:ri}) is the basic model for returns
at the smallest time scale.  We then consider the returns aggregated
on the time scale $\tau$:
\begin{equation} \label{eq:ri_tau}
r_i^\tau(t) = \sum\limits_{\ell=0}^{\tau-1} r_i(t-\ell) ,
\end{equation}
with $t$ being a multiple of $\tau$.  Under the former Gaussian
assumptions, the covariance function of the aggregated returns reads
(see \ref{sec:Acomput}):
\begin{equation}  \label{eq:covar}
C_{ij}^\tau = \langle r_i^\tau(t) r_j^\tau(t) \rangle 
 = \tau \sigma^2 \delta_{ij} + \frac{\beta^2 \Sigma^2 \bigl(\tau(1-\alpha^2) - 2\alpha(1-\alpha^\tau) \bigr)}{(1-\alpha^2)(1-\alpha)^2} \,,
\end{equation}
where $\langle \cdots \rangle$ denotes the expectation, and
$\delta_{ij} = 1$ for $i=j$, and $0$ otherwise.  Note that we set here
$\sigma_i = \sigma$ for all assets for simplicity (this simplification
will be relaxed below).  As we consider the stationary regime, the
covariance function does not depend on time $t$.

Denoting
\begin{equation}  \label{eq:kappa}
\kappa_\alpha(\tau) = \frac{\tau(1-\alpha^2) - 2\alpha(1-\alpha^\tau)}{1-\alpha^2}  \,,
\end{equation}
one gets the correlation matrix
\begin{equation}
\C_{ij}^\tau = \frac{C_{ij}^{\tau}}{\sqrt{C_{ii}^\tau \, C_{jj}^\tau}} = 
\begin{cases} 1 \hskip 11mm i=j ,\cr  \rho^2(\tau) \quad i\ne j, \end{cases}
\end{equation}
with
\begin{equation}  \label{eq:rho_mod}
\rho(\tau) = \bigl(1 + \eta(\tau)/\gamma\bigr)^{-1/2} \,,
\end{equation}
where 
\begin{equation}
\gamma = \frac{\Sigma^2 \beta^2}{\sigma^2}
\end{equation}
and
\begin{equation}  \label{eq:eta}
\eta(\tau) = \frac{\tau}{\kappa_\alpha(\tau)/(1-\alpha)^2} = \frac{(1-\alpha)^2}{1 - \frac{2\alpha}{1-\alpha^2} (1-\alpha^\tau) /\tau} \,.
\end{equation}
The function $\eta(\tau)$, that will play the central role in our
analysis, monotonously decreases from $\eta(1) = 1-\alpha^2$ to
$\eta(\infty) = (1-\alpha)^2$.

Since the matrix $\C^\tau - (1-\rho^2(\tau)) I$ has rank $1$ ($I$
being the identity $N\times N$ matrix), there are $N-1$ eigenvalues
$\lambda_i = 1- \rho^2(\tau)$.  In turn, the single largest eigenvalue
of the correlation matrix $\C^\tau$ can be obtained as follows: $N =
\Tr(\C^\tau) = \lambda_1 + (N-1)\lambda_i$, from which $\lambda_1 = 1
+ (N-1)\rho^2(\tau)$.  We get thus the complete description of the
eigenvalues as functions of the time scale $\tau$:
\begin{eqnarray}  \label{eq:onefactor_lambda}
\lambda_1 &=& 1 + (N-1)\rho^2(\tau)  , \\
\label{eq:onefactor_lambdai}
\lambda_i &=& 1 - \rho^2(\tau)  \hskip 20mm (i=2,3,\ldots,N). 
\end{eqnarray}
In the limit of very large $\tau$, one finds
\begin{equation}
\rho^2(\infty) = \bigl(1 + (1-\alpha)^2/\gamma\bigr)^{-1} .
\end{equation}
This simplest lead-lag one-factor model predicts a monotonous growth
of the largest eigenvalue (corresponding to the market mode) with the
time scale $\tau$, up to a saturation plateau.  In turn, the other
eigenvalues exhibit a monotonous decrease to a plateau.  In spite of
the exponential decay of the lead-lag memory effect in
Eq. (\ref{eq:ri}), the approach to the plateau is governed by a slow,
$1/\tau$ power law, in a qualitative agreement with the empirical
observation (see Sec. \ref{sec:application} for quantitative
comparison).  In particular, this approach has no well-defined time
scale.

While the basic model can potentially capture the behavior of the
largest eigenvalue, it clearly fails to distinguish other eigenvalues.
One needs therefore to relax some simplifying assumptions to render
the model more realistic.

\subsection{General lead-lag one-factor model}

We start by introducing arbitrary volatilities $\sigma_i$ and
sensitivities $\beta_i$ of the $i$-th asset to the common factor
$R(t)$:
\begin{equation}
r_i(t) = \ve_i(t) + \beta_i \sum\limits_{k=0}^\infty \alpha^k R(t-k) .
\end{equation}
In this case, the computation is precisely the same, the only
difference is that
\begin{equation}
C_{ij}^\tau = \tau \sigma_i^2 \delta_{ij} + \Sigma^2 \beta_i \beta_j \kappa_\alpha(\tau) .
\end{equation}
As a consequence, the structure of the correlation matrix is fully
determined by $\beta_i$, whereas the dependence on the time scale
$\tau$ is still represented by $\kappa_\alpha(\tau)$.  The correlation
matrix reads
\begin{equation}  \label{eq:rhoij_tau}
\C_{ij}^\tau = \begin{cases} 1  \hskip 20mm (i=j) ,\cr \rho_i(\tau) \rho_j(\tau)  \quad (i\ne j), \end{cases}
\end{equation}
with
\begin{equation}  \label{eq:xii}
\rho_i(\tau) = \bigl(1 + \eta(\tau)/\gamma_i\bigr)^{-1/2} \,, \qquad \gamma_i = \frac{\Sigma^2 \beta_i^2}{\sigma_i^2} \,.
\end{equation}
The eigenvalues of this correlation matrix can be computed as follows.

If all $\gamma_i$ are distinct%
\footnote{
When some $\gamma_i$ are identical, the analysis of eigenvalues
becomes more involved (see \ref{sec:new_model}), but the largest
eigenvalue still satisfies Eq. (\ref{eq:lambda_eq_xi}) and can thus be
approximated by Eq. (\ref{eq:lambda1_app}).  In particular, if all
$\gamma_i = \gamma$, one gets $\lambda_1 \approx N \rho^2(\tau)$,
which is close to the exact solution (\ref{eq:onefactor_lambda}).},
%footnote
the components of an eigenvector are
\begin{equation}
v_i = \frac{\rho_i Q}{\lambda-1+\rho_i^2} \quad (i=1,\ldots,N), \qquad \textrm{with} ~~ Q = \sum\limits_{i=1}^N \rho_i v_i \,,
\end{equation}
from which one gets the equation on the eigenvalues $\lambda$
\begin{equation}  \label{eq:lambda_eq_xi}
\sum\limits_{i=1}^N \frac{\rho_i^2}{\lambda-1+\rho_i^2} = 1 .
\end{equation}
This equation has $N$ distinct solutions that can be characterized in
terms of $\rho_i^2$ (see \ref{sec:new_model}).  When $N$ is large, the
largest eigenvalue is expected to be large, and the asymptotic
expansion of Eq. (\ref{eq:lambda_eq_xi}) yields
\begin{equation}  \label{eq:lambda1_app}
\lambda_1 \approx \sum\limits_{i=1}^N \rho_i^2 = \sum\limits_{i=1}^N \bigl(1 + \eta(\tau)/\gamma_i \bigr)^{-1} \,.
\end{equation}
In turn, the other eigenvalues are below $1$ (see
\ref{sec:new_model}).  As a consequence, such a lead-lag one-factor
model cannot reproduce several eigenvalues larger than $1$.  For this
purpose, one needs to consider multiple factors.

\subsection{General lead-lag multi-factor model}
\label{sec:extensions}

Now we consider a general lead-lag multi-factor model
\begin{equation}  \label{eq:model_multi-factor}
r_i(t) = \ve_i(t) + \sum\limits_{k=0}^\infty \alpha^k \sum\limits_{f=1}^F \beta_{i,f} \, R_f(t-k) ,
\end{equation}
where $\ve_i(t)$ are independent centered Gaussian variables
(representing random fluctuations specific to the stock $i$) with
variance $\sigma_i^2$, $F$ is the number of factors, $R_f(t)$ are
independent centered Gaussian returns of the factor $f$ with variance
$\Sigma_f^2$, $\beta_{i,f}$ is the sensitivity of the stock $i$ to the
factor $f$, and $\alpha$ sets the relaxation time.  Repeating the
computation from \ref{sec:Acomput}, one gets
\begin{equation}  \label{eq:Ctau_two}
\C_{ij}^\tau = \delta_{ij} + (1-\delta_{ij}) \sum\limits_{f=1}^F \rho_{i,f} \rho_{j,f} ,
\end{equation}
where
\begin{equation}  \label{eq:rho_12}
\rho_{i,f}(\tau) = \frac{\Sigma_f \beta_{i,f}}{\beta_i} \, \rho_i(\tau) \,
\end{equation}
and
\begin{equation}  \label{eq:prop1}
\rho_i(\tau) = \bigl(1 +  \eta(\tau)/\gamma_i \bigr)^{-1/2} \,,  \qquad
\gamma_i = \frac{\beta_i^2}{\sigma_i^2} \,, \qquad
\beta_i^2 = \sum\limits_{f=1}^F \Sigma_f^2 \, \beta_{i,f}^2 \,.
\end{equation}
Considering $\rho_{i,f}$ as the elements of an $N\times F$ matrix
$\rho$, one can rewrite Eq. (\ref{eq:Ctau_two}) in a matrix form
\begin{equation}
\C^\tau = (I - P) + \rho \rho^\dagger ,
\end{equation}
where $P$ is the diagonal matrix formed by $\rho_i^2$, and $\dagger$
denotes the matrix transpose.  

The matrix $\rho$ of size $N\times F$ plays the central role in the
following analysis.  As the elements of the matrix $\rho$ are real,
$\rho \rho^\dagger$, as well as $\rho^\dagger \rho$, are positive
semi-definite matrices which have nonnegative eigenvalues.  The rank
of the matrix $\rho$ is equal to that of matrices $\rho \rho^\dagger$
and $\rho^\dagger \rho$ and thus cannot exceed $\min\{F,N\}$.  Given
that $F \ll N$, the correlation matrix $\C^\tau$ appears as the
perturbation of a diagonal matrix by a low-rank matrix.

The eigenvalues of the correlation matrix are the zeros of the
determinant
\begin{equation}  \label{eq:det_lambda1}
0 = \det\bigl(\lambda I - \C^\tau\bigr) = \det\bigl(\lambda I - I + P - \rho\rho^\dagger\bigr) .
\end{equation}
Since $\rho \rho^\dagger$ is a low-rank perturbation, one can expect,
as in the one-factor case of \ref{sec:new_model}, that most
eigenvalues coincide with that of the unperturbed diagonal matrix
$I-P$, i.e., they are given by $1-\rho_i^2$ for some indices $i$.
These eigenvalues are essentially hidden by noise and non-exploitable
in practice.  We are interested in large eigenvalues that
(significantly) exceed $1$.

If $\lambda$ exceeds $1$, it cannot be equal to $1-\rho_i^2$ for all
$i$, the matrix $\lambda I - I + P$ is nonsingular, its inverse
exists, so that one can rewrite Eq. (\ref{eq:det_lambda1}) as
\begin{equation}
0 = \det\bigl(\lambda I - I + P\bigr) \, \det\bigl(I - \rho^\dagger (\lambda I - I + P)^{-1} \rho\bigr) ,
\end{equation}
from which one gets a new equation on eigenvalues:
\begin{equation}  \label{eq:det_lambda2}
0 = \det\bigl(I - \underbrace{\rho^\dagger (\lambda I - I + P)^{-1} \rho}_{\phi(\lambda)} \bigr) .
\end{equation}
(here we used a general property: if $A\in \CC^{m\times m}$ is
nonsingular matrix and $U,V \in \CC^{m\times r}$, then $\det(A + UV^*)
= \det(A) \det(I + V^* A^{-1} U)$, see \cite{Xiong09}).  Denoting the
$F\times F$ matrix in the determinant as $\phi(\lambda)$, one can
write explicitly its elements as
\begin{equation}
\phi_{f,g}(\lambda) = \sum\limits_{i=1}^N \frac{\rho_{i,f} \, \rho_{i,g}}{\lambda - 1 + \rho_i^2} \,.
\end{equation}
The solutions of Eq. (\ref{eq:det_lambda2}) determine some eigenvalues
$\lambda$ of the correlation matrix in Eq. (\ref{eq:Ctau_two}).  As
one typically deals with the situation $N \gg F$, the reduction of the
original determinant equation (\ref{eq:det_lambda1}) for a matrix of
size $N\times N$ to Eq. (\ref{eq:det_lambda2}) for a matrix of size
$F\times F$ is a significant numerical simplification of the problem.
Most importantly, this formal solution allows one to get analytical
insights onto the eigenvalues, as we did in the one-factor case in
\ref{sec:new_model}.
Note that in the one-factor case ($F = 1$), the determinant equation
(\ref{eq:det_lambda2}) is simply reduced to
\begin{equation}  \label{eq:det_lambda_F1}
0 = \det(I - \phi(\lambda)) = 1 - \phi_{1,1}(\lambda) = 1 - \sum\limits_{i=1}^N \frac{\rho_i^2}{\lambda - 1 + \rho_i^2} \,,
\end{equation}
i.e., we retrieve Eq. (\ref{eq:lambda_eq_xi}).

If one searches for large eigenvalues, $\lambda \gg 1$, one can
neglect the matrix $P-I$ in comparison to $\lambda I$ in
Eq. (\ref{eq:det_lambda2}), that yields
\begin{equation}  \label{eq:det_lambda3}
\det\bigl(\lambda I - \rho^\dagger \rho\bigr) = 0 .
\end{equation}
In other words, the large eigenvalues of the correlation matrix can be
approximated by the eigenvalues of the matrix $\rho^\dagger \rho$ of
size $F\times F$.  This symmetric positive semi-definite matrix has
$F$ nonnegative eigenvalues that correspond to $F$ factors.

\subsection{Practical approximation}

As we will discuss in detail in Sec. \ref{sec:application}, empirical
data exhibit the short-range memory effect ($\alpha$ is small) and the
relatively small impact of the factors onto the variance of individual
stocks as compared to the stock-specific fluctuations ($\gamma_i$ are
small).  In this situation, which is particular to the time series of
securities returns at the considered time scales, one has
$\eta(\tau)/\gamma_i \gg 1$ so that $\rho_i(\tau)$ in
Eq. (\ref{eq:prop1}) can be approximated as 
\begin{equation}
\rho_i^2(\tau) \simeq \frac{\gamma_i}{\eta(\tau)}.  
\end{equation}
This approximation greatly simplifies the elements of the matrix
$\rho^\dagger \rho$:
\begin{equation}
(\rho^\dagger \rho)_{f,g} = \sum\limits_{i=1}^N \underbrace{\frac{\Sigma_f \beta_{i,f} \rho_i(\tau)}{\beta_i}}_{= \rho_{i,f}}
 \, \underbrace{\frac{\Sigma_g \beta_{i,g} \rho_i(\tau)}{\beta_i}}_{= \rho_{i,g}}
\approx \frac{N}{\eta(\tau)} \,  \Gamma_{f,g} \,,
\end{equation}
where the matrix elements $\Gamma_{f,g}$ do not depend on the time scale:
\begin{equation}
\Gamma_{f,g} = \frac{\Sigma_f \Sigma_g}{N} \sum\limits_{i=1}^N \frac{\beta_{i,f} \beta_{i,g}}{\sigma_i^2} \,.
\end{equation}
As a consequence, all the elements of the matrix $\rho^\dagger \rho$
and thus its eigenvalues exhibit the same dependence on the time scale
$\tau$, expressed via the explicit function $\eta(\tau)$ given by
Eq. (\ref{eq:eta}).  Denoting the eigenvalues of the matrix $\Gamma$
as $\gamma_f$ ($f = 1,\ldots,F$), one gets the following approximation
for large eigenvalues of the correlation matrix:
\begin{equation}  \label{eq:main_app}
\lambda_f \approx \frac{N \gamma_f}{\eta(\tau)} \qquad (f=1,\ldots,F)\,.
\end{equation}
From the explicit form (\ref{eq:xii}) of $\eta(\tau)$, one deduces a
slow, $1/\tau$, power law approach of the eigenvalue to the saturation
level as the time scale $\tau$ increases.  Within this approximate
computation, all large eigenvalues exhibit the same dependence on the
time scale.

In practice, one aims at constructing the factors $R_f$ to capture
independent features of cross-correlations in the market.  The
sensitivies $\beta_{i,f}$ and $\beta_{i,g}$ of the stock $i$ to
factors $R_f$ and $R_g$ are thus expected to be ``orthogonal'', and
this property can be formally expressed by requiring that the
nondiagonal elements of the matrix $\Gamma$ are negligible.  In this
case, the eigenvalues $\gamma_f$ are given by the diagonal elements
\begin{equation}
\gamma_f = \Gamma_{f,f} = \frac{1}{N} \sum\limits_{i=1}^N \frac{\Sigma_f^2 \beta_{i,f}^2}{\sigma_i^2} \,.
\end{equation}
This is a kind of empirical mean of the squared sensitivities
$\beta_{i,f}^2$, normalized by the squared volatilities $\sigma_i^2$.

\subsection{Application to empirical data}
\label{sec:application}

\begin{table}
\begin{center}
\begin{tabular}{c c c c c }  \hline
$f$              &   1  &    2 &   3  &   4  \\  \hline
$\gamma$         & 0.17 & 0.03 & 0.02 & 0.01 \\
$\alpha$         & 0.16 & 0.25 & 0.18 & 0.26 \\
$t_\alpha$ (min) & 0.55 & 0.72 & 0.58 & 0.74 \\  \hline
\end{tabular}
\end{center}
\caption{
Two adjustable parameters of the fitting formula (\ref{eq:main_app})
applied to four largest eigenvalues of the correlation matrix of $N =
533$ U.S. stocks' returns.  The corresponding relaxation time
$t_\alpha$ in minutes is obtained as $1~{\rm min}/\ln(1/\alpha)$.}
\label{tab:fit}
\end{table}
%   5   104.8998    0.1889

We aim at applying the lead-lag factor model to fit the eigenvalues of
the empirical correlation matrix of U.S. stocks' returns.  The fitting
formula (\ref{eq:main_app}) has two adjustable parameters: the
relaxation time $\alpha$ in the function $\eta(\tau)$ and the
amplitude $N\gamma_f$.  Using the least square fitting algorithm
implemented as the routine \verb|lsqcurvefit| in Matlab, we apply the
formula (\ref{eq:main_app}) separately to each empirical eigenvalue.

\begin{figure}
\begin{center}
\includegraphics[width=65mm]{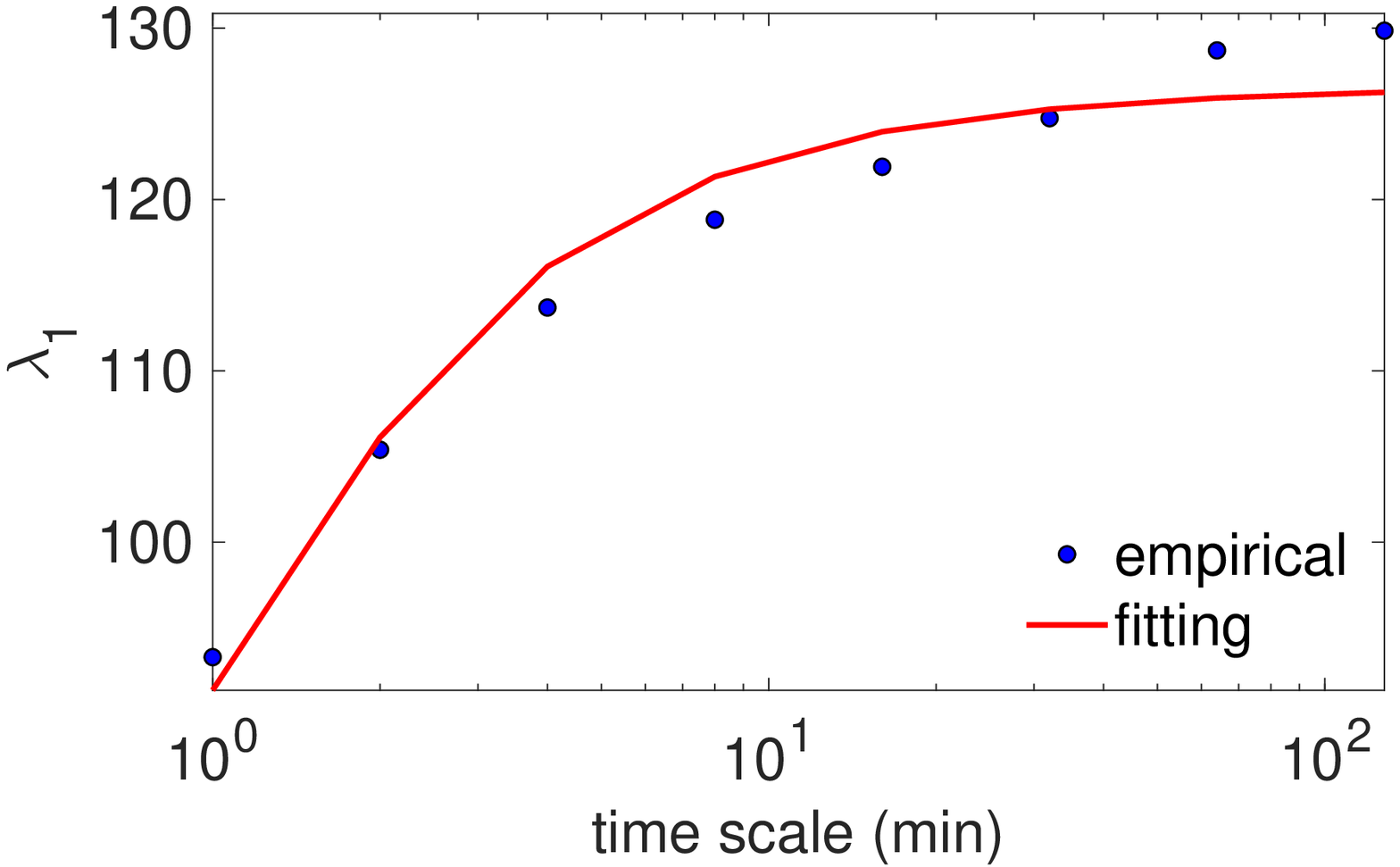} % USA_rho_eigen1_fit3.eps}
\includegraphics[width=65mm]{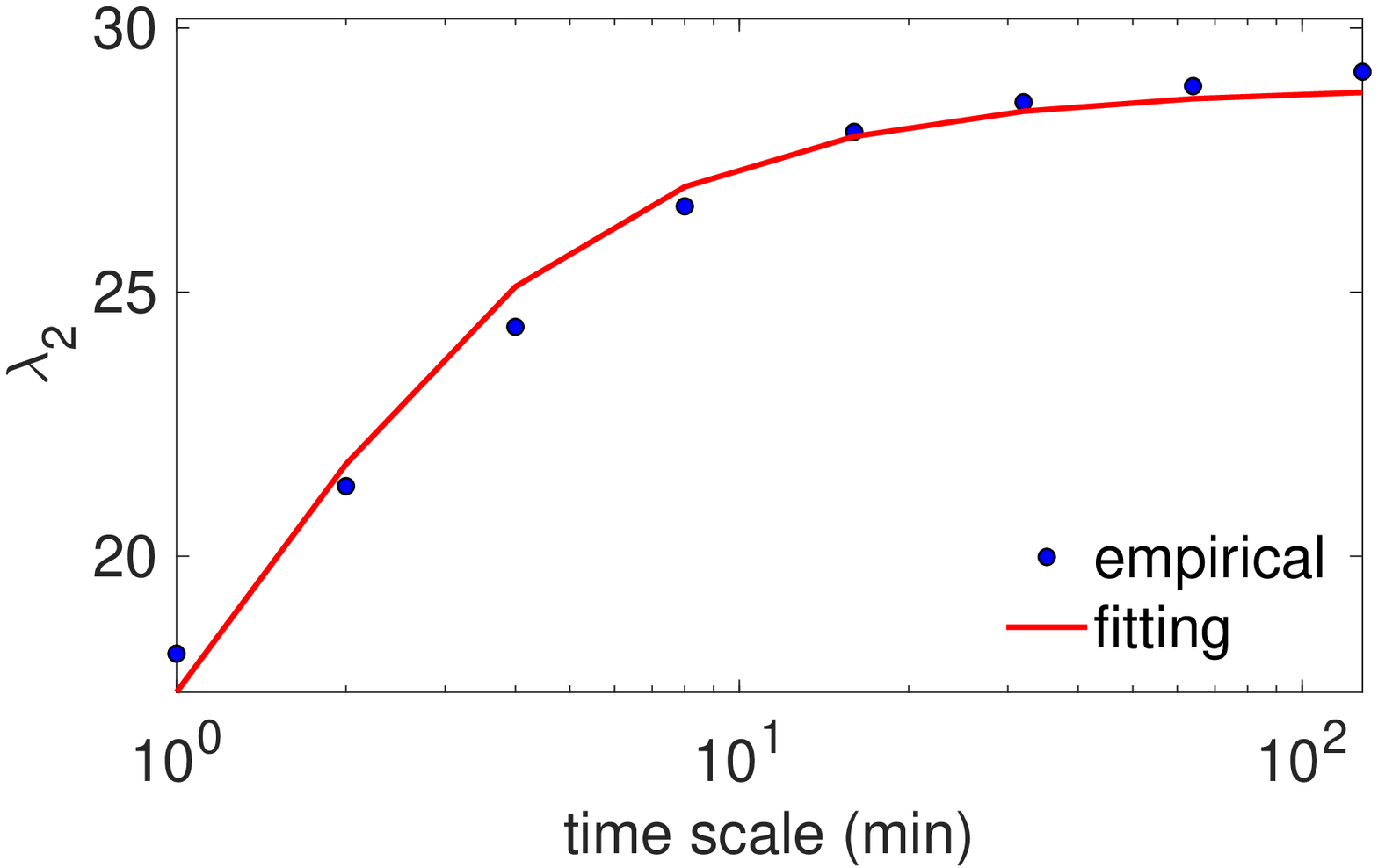} % USA_rho_eigen2_fit3.eps}
\includegraphics[width=65mm]{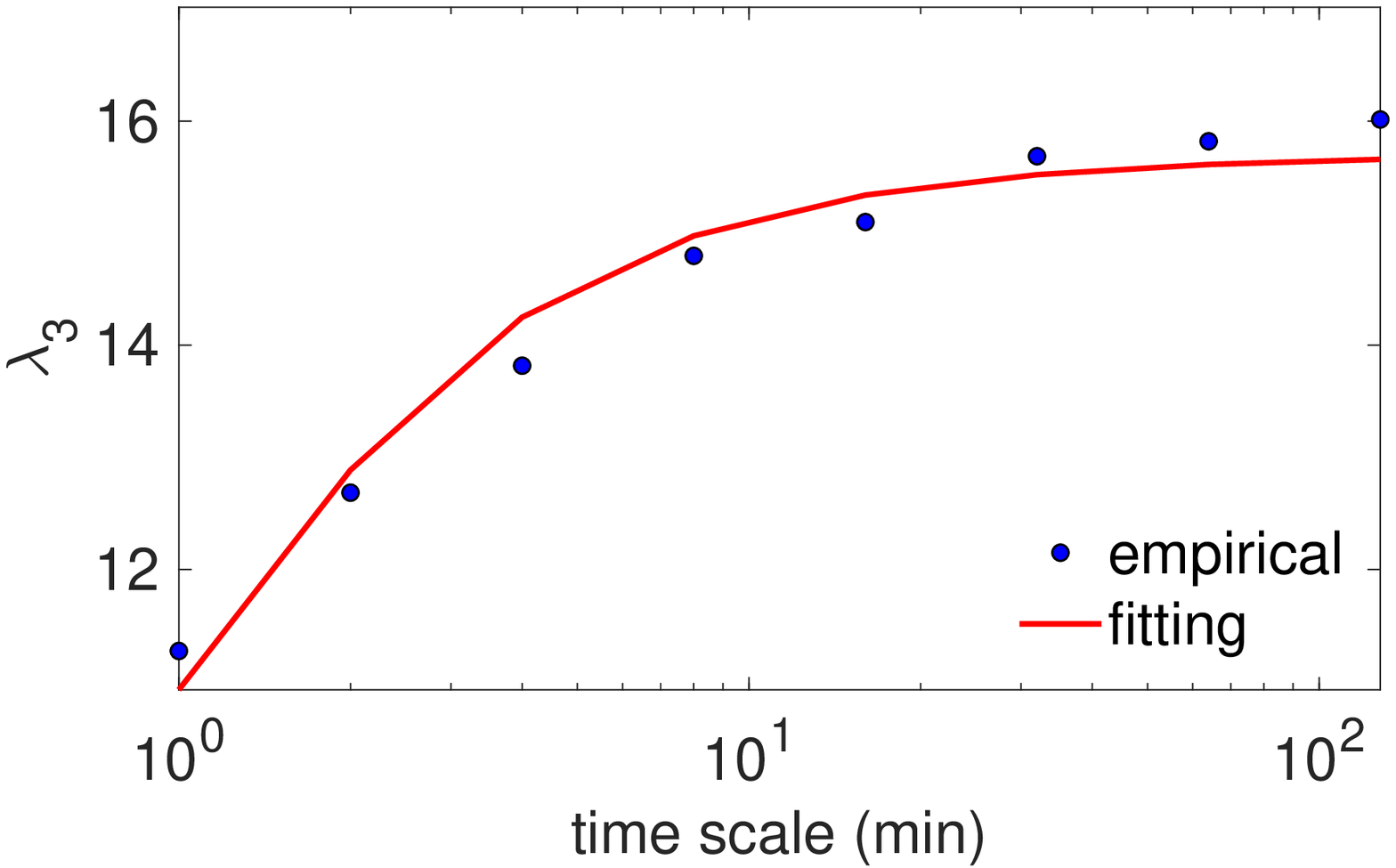} % USA_rho_eigen3_fit3.eps}
\includegraphics[width=65mm]{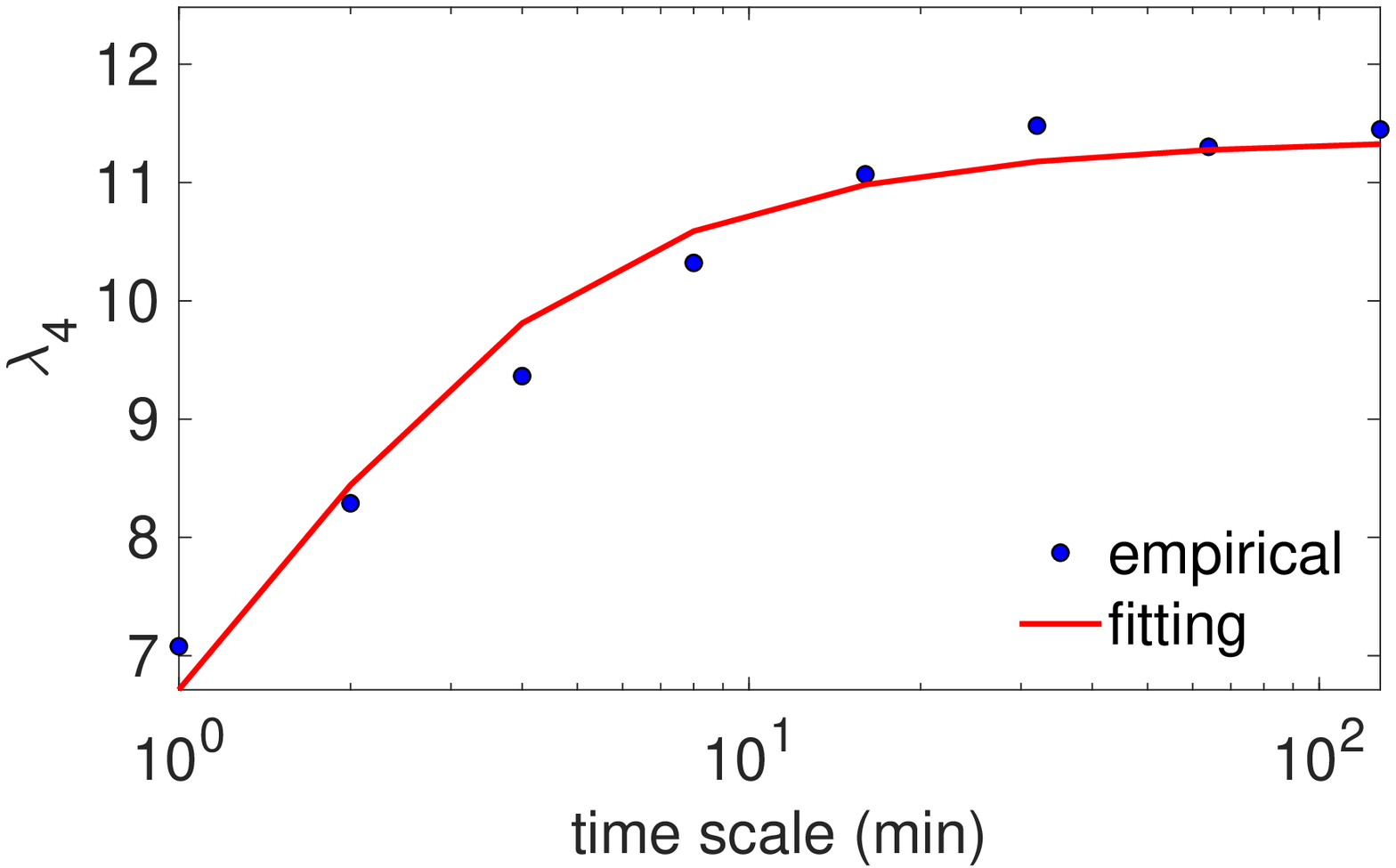} % USA_rho_eigen4_fit3.eps}
\end{center}
\caption{
Fitting by Eq. (\ref{eq:main_app}) of the four largest eigenvalues of
the correlation matrix of $N = 533$ U.S. stocks' returns, computed by
aggregating 1-min returns with the time scale $\tau$.  The adjustable
parameters $\alpha$ and $\gamma$ are summarized in Table
\ref{tab:fit}. }
\label{fig:USA_rho_eigen_fit}
% [LamC, Lamrho, Cdiag] = JLI_intraday_create_C_load(0);
% [LamC, Lamrho, x,yx] = JLI_intraday_eigen_fit_fig2(2, LamC, Lamrho);
\end{figure}

Figure \ref{fig:USA_rho_eigen_fit} shows the fitting of the four
largest eigenvalues.  The good quality of the fit by the lead-lag
factor model indicates that, in spite of numerous simplifying
assumptions on which the model was built, it captures the overall
behavior qualitatively well.  In particular, the eigenvalues converge
to limiting values, at least for the considered short-time scales (up
to 2 hours).  Moreover, this saturation level is approached slowly,
with the characteristic $1/\tau$ power law dependence.  The adjustable
parameters are summarized in Table \ref{tab:fit}.  Rewriting the
attenuation factor $\alpha^k$ in the lead-lag factor model
(\ref{eq:ri}) as $\exp(-t/t_\alpha)$ with $t = k \tau_0$ and $t_\alpha
= \tau_0/\ln (1/\alpha)$, where $\tau_0 = 1$~min is the finest time
scale of the time series used, one gets the relaxation time $t_\alpha$
in minutes.  One can see that the relaxation times $\alpha$ (or
$t_\alpha$) for four eigenvalues are close to each other.  In other
words, all the dominant eigenmodes evolve at comparable time scales.
This is an important conclusion which refutes a common belief that the
market mode (corresponding to the largest eigenvalue) evolves at a
time scale that is significantly different from other modes (sectors
and style factors).  The values of $t_\alpha$ are of the order of one
minute, in agreement with predictions by Benzaquen {\it et al.}
\cite{Benzaquen17}.  Remarkably, while the lead-lag memory effects
vanish so rapidly, they impact the behavior of the eigenvalues at much
longer time scales.  In particular, if the lead-lag was ignored (by
setting $\alpha = 0$), the largest eigenvalue would be $\simeq
N\gamma_1$ and independent of the time scale $\tau$.  For instance,
using the estimated value $\gamma_1 = 0.17$ and setting $\alpha = 0$,
one would get the largest eigenvalue to be $90$, which is
significantly smaller than the expected limit $128$ for $\alpha =
0.16$ or the observed value $130$ at $\tau = 128$~min.

\section{Conclusion}
\label{sec:conclusion}

We investigated the dependence of the eigenvalues of the correlation
matrix on the time scale $\tau$.  Aggregating 1-min returns of the
largest 533 U.S. stocks (2013-2017) to estimate the correlation matrix
at different time scales, we showed that its large eigenvalues grow
with $\tau$ and apparently saturate to limiting values.  This growth
reflects the important phenomenon that inter-stock correlations
accumulate over time scales.

To rationalize this phenomenon and to interpret empirical
observations, we developed the lead-lag factor model.  In the
one-factor case, each stock is considered to be partly correlated to a
given lead-lag factor.  Under several simplifying assumptions, we
derived a simple formula for large relevant eigenvalues.  This formula
containing just two easily interpretable adjustable parameters, was
then validated on empirical data.

The relaxation time of the stock market was estimated to be around 1
minute.  A possible interpretion of this observation can be that a
transaction can generate a cascade of transactions that decays in 1
minute so that the impact of transaction on price decays in 1 minute.
As correlations emerge from the cross-impact of transactions on
prices, we model this effect by extending the Kyle model to the impact
of transaction on preferential portfolios with a lead lag effect.

The small value of the observed relaxation time suggests that
correlation measurements based on 5 minutes returns should provide a
good proxy of correlation of daily returns for risk management, in
line with the conclusion by Liu {\it et al.} on volatility estimation
\cite{Liu15}.  However, other phenomena are likely to occur at much
larger time scales (from day to month), e.g., autocorrelations of
returns of financial factors (book, size, momentum) due to herding
effect, or lack of liquidity.  An accurate estimation of correlations
at larger time scales remains a challenging problem because of a
limited number of the available returns and thus higher impact of
noise in the estimated correlation matrix.  To overcome this
limitation, one can either consider time horizons over several decades
(in which case neglecting variations of corrections over time becomes
debatable), or reduce the number of considered securities and thus the
dimension of the correlation matrix (in which case financial meaning
of estimated correlations may be debatable).  A possible solution
consists in constructing relevant financial factors and investigating
how their correlations change with the time scale, as suggested by our
factor-based model.

%%%%%%%%%%%%%%%%%%%%%%%%%%%%%%%%%%%%%%%%%%%%%%%%%%%%%%%%%%%%%%
\appendix
\section{Computation of the covariance matrix}
\label{sec:Acomput}

The covariance matrix of aggregated centered Gaussian returns
$r_i^\tau(t)$ defined by Eq. (\ref{eq:ri_tau}) is
\begin{eqnarray}
C_{ij}^\tau &=& \langle r_i^\tau(t) r_j^\tau(t) \rangle \\  \nonumber
&=& \tau \sigma^2 \delta_{ij} 
+ \beta^2 \sum\limits_{\ell_1,\ell_2 =0}^{\tau-1} \sum\limits_{k_1,k_2=0}^\infty \alpha^{k_1} \alpha^{k_2}  
\langle R(t-\ell_1-k_1) R(t-\ell_2-k_2)\rangle . 
\end{eqnarray}
The first term in this expression comes from the uncorrelated
stock-dependent fluctuations.  The independence of returns $R(k)$
implies
\begin{equation}
C_{ij}^\tau = \tau \sigma^2 \delta_{ij} 
+ \beta^2 \sigma_m^2 \sum\limits_{\ell_1,\ell_2 =0}^{\tau-1} \sum\limits_{k_1,k_2=0}^\infty \alpha^{k_1} \alpha^{k_2}  
\delta_{\ell_1+k_1,\ell_2+k_2} .
\end{equation}
To calculate these four sums, it is convenient to consider separately
various terms depending on $\ell_1$ and $\ell_2$:

$\bullet$ there are $\tau$ terms with $\ell_1 = \ell_2$ that implies
$k_1 = k_2$, whose contribution is
\begin{equation}
\tau \sum\limits_{k=0}^\infty \alpha^{2k} = \frac{\tau}{1-\alpha^2} \,;
\end{equation}

$\bullet$ there are $\tau-1$ terms with $\ell_1 = \ell_2 + 1$ that
implies $k_1 = k_2 - 1$, whose contribution is
\begin{equation}
(\tau-1) \sum\limits_{k=0}^\infty \alpha^{2k+1} = \frac{(\tau-1)\alpha}{1-\alpha^2} \,.
\end{equation}
Moreover, the same contribution comes from $\ell_1 = \ell_2 - 1$ and
$k_1 = k_2 + 1$.

$\bullet$ similarly, there are $\tau-j$ terms with $\ell_1 = \ell_2 +
j$ that implies $k_1 = k_2 - j$, whose contribution is
\begin{equation}
(\tau-j) \sum\limits_{k=0}^\infty \alpha^{2k+j} = \frac{(\tau-j)\alpha^j}{1-\alpha^2} \,,
\end{equation}
and this contribution is doubled by the symmetry argument.

$\bullet$ finally, there is one term with $\ell_1 = \ell_2 + (\tau-1)$
and thus $k_1 = k_2 - (\tau-1)$ whose contribution is
$\alpha^{\tau-1}/(1-\alpha^2)$.

Combining all these terms, one gets after simplifications
Eq. (\ref{eq:covar}).

\section{Analysis of the lead-lag one-factor model}
\label{sec:new_model}

We study in more detail the model (\ref{eq:rhoij_tau}) of the
correlation matrix $\C$, with $\rho_i(\tau)$ given by
Eq. (\ref{eq:xii}).  This matrix is a perturbation of the identity
matrix by a rank one matrix, for which many spectral properties are
known (see, e.g., \cite{Golub73}).
This matrix combines both effects: the correlation coefficient $\rho$
and the impact of the exponential moving average (with the coefficient
$\alpha$).  We search for an eigenvector of this matrix as $v =
(v_1,v_2,\ldots,v_n)^\dagger$.  Writing explicitly $\C v = \lambda v$,
we get
\begin{equation}  \label{eq:vi_eq}
v_i (1 - \rho_i^2) + \rho_i Q = \lambda v_i  \qquad (i=1,\ldots,N),
\end{equation}
where
\begin{equation} \label{eq:Q_aux}
Q = \sum\limits_{i=1}^N v_i \rho_i .
\end{equation}

First, we note that if $\rho_i = 0$ for some $i$, then the above
equation is reduced to $v_i = \lambda v_i$ that has two solutions:
either $\lambda = 1$ and $v_i$ can be arbitrary; or $v_i = 0$.  One
can check that if $\rho_{i_1} = \ldots = \rho_{i_k} = 0$ for $k$
stocks, then the correlation matrix has the eigenvalue $\lambda = 1$
with the multiplicity $k$.  The corresponding eigenvectors can be
chosen as an orthogonal basis in the subspace $\R^k$.  In turn, the
remaining $n-k$ eigenvalues are nontrivial, and can be determined as
discussed below.  In what follows, we focus on these nontrivial
eigenvalues, i.e., we assume that all $\rho_i \ne 0$.

The equation (\ref{eq:vi_eq}) has two solutions: 

(i) either $\lambda = 1 - \rho_i^2$ and $Q = 0$; or 

(ii) $\lambda \ne 1 - \rho_i^2$ and 
\begin{equation}
v_i = \frac{\rho_i Q}{\lambda - 1 + \rho_i^2} \,.
\end{equation}
In the latter case, one can substitute this expression into
Eq. (\ref{eq:Q_aux}) to get an equation on the eigenvalue $\lambda$:
\begin{equation}
\label{eq:lambda_eq}
\sum\limits_{i=1}^N \frac{\rho_i^2}{\lambda - 1 + \rho_i^2} = 1 .
\end{equation}
This equation can be seen as a polynomial of degree $N$ which has $N$
({\it a priori} complex-valued) zeros.  Finally, $Q$ can be fixed by
setting the normalization condition on $v$:
\begin{equation}
1 = \sum\limits_{i=1}^N v_i^2 = Q^2 \sum\limits_{i=1}^N \frac{\rho_i^2}{(\lambda-1+\rho_i^2)^2} \, .
\end{equation}
This is a generic situation.

Let us return to the first option, namely, we suppose that $\lambda =
1 - \rho_k^2$ for some index $k$ that implies that $Q = 0$.  If all
$\rho_i$ are distinct, i.e., $\rho_1 \ne \rho_2 \ne \ldots \ne
\rho_N$, so that $v_i = 0$ for all $i\ne k$, but, due to $Q = 0$, it
would also imply that $v_k = 0$.  As a consequence, $v = 0$ but this
is not an eigenvector.  We conclude that, if all $\rho_i$ are
distinct, then $\lambda$ cannot be given by $1-\rho_i^2$, and this
option is excluded.

Now, we consider the case when two or more values $\rho_i$ are
identical.  For instance, let us assume that $\rho_1 = \rho_2 \ne
\rho_3 \ne \ldots \ne \rho_N$.  In this case, $\lambda = 1 - \rho_1^2$
is indeed an eigenvalue.  In fact, one gets $Q = 0$ and thus $v_i = 0$
for $i > 2$.  However, one has $Q = \rho_1 v_1 + \rho_2 v_2 =
\rho_1(v_1 + v_2) = 0$, implying that $v_1 = - v_2$.  The
normalization condition implies thus $v_1 = - v_2 = 1/\sqrt{2}$.  We
conclude that $\lambda = 1 - \rho_1^2$ is then a single eigenvalue.
More generally, if $\rho_1 = \rho_2 = \ldots = \rho_k \ne \rho_{k+1}
\ne \ldots \ne \rho_N$, then the eigenvalue $\lambda = 1-\rho_1^2$ has
the multiplicity $k-1$.

In general, it is convenient to denote $z_i = 1 -\rho_i^2$ and to
order them in an increasing order:
\begin{equation}
z_1 \leq z_2 \leq z_3 \leq \ldots \leq z_N
\end{equation}
or, equivalently, by grouping the eventual identical values:
\begin{equation}
\begin{split}
z_1 & = z_2 = \ldots = z_{i_1} < z_{i_1+1} = z_{i_1+2} = \ldots = z_{i_1+i_2} \\
& < \ldots < z_{i_1+\ldots + i_m} = z_{i_1 + \ldots + i_m+1} = \ldots = z_N .  \\
\end{split}
\end{equation}
In other words, there are $i_1$ identical values $z_1 = \ldots =
z_{i_1}$; $i_2$ identical values $z_{i_1+1} = \ldots = z_{i_1+i_2}$,
etc. (note that when all $z_i$ are distinct, one has $i_1 = i_2 =
\ldots = 1$).  In this configuration, the correlation matrix has: the
eigenvalue $z_1$ with the multiplicity $i_1-1$ (if $i_1 > 1$); the
eigenvalue $z_2$ with the multiplicity $i_2-1$ (if $i_2 > 1$); etc.
If for some $k$, $z_{i_k} = 1$, then this eigenvalues has the
multiplicity $i_k$.  Finally, the remaining eigenvalues are determined
as solutions of Eq. (\ref{eq:lambda_eq}) that can be written as $f(z)
= 1$, with
\begin{equation}
f(z) = \sum\limits_{i=1}^N \frac{\rho_i^2}{z - 1 + \rho_i^2} = \sum\limits_{i=1}^N \frac{1-z_i}{z - z_i}  \,.
\end{equation}
The terms with $z_i = 1$ (resulting in the eigenvalue $\lambda = 1$)
are excluded from this sum.  Moreover, if some $z_i$ are identical,
the corresponding terms are just grouped together.  As a consequence,
the equation $f(z) = 1$ is reduced to a polynomial of degree at most
$N$ (the degree $N$ corresponding to the case when all $z_i$ are
distinct).

It is worth noting that the function $f(z)$ is decreasing everywhere:
\begin{equation}
f'(z) = - \sum\limits_{i=1}^N \frac{\rho_i^2}{(z-z_i)^2} < 0 .
\end{equation}
As a consequence, one gets immediately that each interval
$(z_i,z_{i+1})$ (with $z_i < z_{i+1}$ and $z_{N+1} = \infty$) has
exactly one solution of the equation $f(z) = 0$, i.e., one eigenvalue.
In particular, one gets the following bounds for the smallest
eigenvalue
\begin{equation}
z_1 \leq \min\limits_{1\leq i\leq N} \{\lambda_i\} \leq z_2 .
\end{equation}
We conclude that all eigenvalues are positive if and only if $z_1
\geq 0$, i.e., $\rho_i^2 \leq 1$ for all $i$.  In other words, the
inequalities $\rho_i^2 \leq 1$ for all $i$ present the necessary and
sufficient condition for the positive definiteness of the matrix.
These conditions are evidently satisfied in our setting.

Since $f(1) \geq 1$, one also gets the following bound for the largest
eigenvalue
\begin{equation}
\lambda_1 = \max\limits_{1\leq i\leq N} \{\lambda_i\} \geq 1 
\end{equation}
(note that the eigenvalues are ordered in descending order, $\lambda_1
\geq \lambda_2 \geq \ldots$, in contrast to $z_k$).  However, this
bound is rather weak.  In turn, since $\lambda_{2}
\leq z_N = 1 - \rho_N^2 < 1$, all other eigenvalues are below $1$:
\begin{equation}
\lambda_i < 1  \qquad (i=2,3,\ldots,N).
\end{equation}

%%%%%%%%%%%%%%%%%%%%%%%%%%%%%%%%%%%%%%%%%%%%%%%%%%%%%%%%%%%%%%%%%%%%%%%%%%%%%%%%%%
%\newpage

\end{document}